\documentstyle[12pt]{article}
\renewcommand{\theequation}%
{\thesection.\arabic{equation}}
\begin{document}
\begin{center}
{\Large {\bf Integrals of periodic motion and periodic solutions
for classical equations of relativistic string with masses at ends \\}}
\bigskip
{\large {\bf I. Integrals of periodic motion \\}}
\end{center}
\begin{center}
{\large B.M.Barbashov \footnote{e-mail address: 
barbash\verb|@|thsun1.jinr.dubna.su}\\}
Bogoliubov Laboratory of Theoretical Physics \\ JINR, 141980, Dubna,
Russia
\end{center}
\begin{center}
\parbox[c]{120mm}
{\small Boundary equations for the relativistic string with masses at 
ends are formulated in terms of geometrical invariants of world 
trajectories of masses at the string ends. In the three--dimensional 
Minkowski space $E^1_2$, there are two invariants of that sort, the 
curvature $K$ and torsion $\kappa$.  Curvatures of trajectories of 
the string ends with masses are always constant, $K_i = \gamma/m_i (i 
=1,2,)$, whereas torsions $\kappa_i(\tau)$ obey a system of 
differential equations with deviating arguments. For these equations 
with periodic $\kappa_i(\tau+n l)=\kappa(\tau)$, constants of motion 
are obtained (part I) and exact solutions are presented (part II) for 
periods $l$ and $2l$ where $l$ is the string length in the plane of 
parameters $\tau$ and $\sigma \  (\sigma_1 = 0, \sigma_2 =l)$.} 
\end{center}

\section*{Introduction} The relativistic string with point masses at 
ends is the dynamic basis of the string model of hadrons since there 
is a direct analogy between an open string with masses at its ends 
and a quark--antiquark pair connected by a tube of the gluon field in 
quantum chromodynamics [1]. Difficulties of the hadron string model 
are due to the nonlinear character of boundary conditions, and even 
at the classical level, the investigation of this system becomes a 
complicated mathematical problem whose general solution is not yet 
derived.  In ref.[2], boundary equations for the relativistic string 
in the three-- dimensional Minkowski space $E_2^1(t,x,y)$ have been 
formulated in terms of geometrical invariants of world trajectories 
of the string ends with masses, their constant curvature $K$ and 
torsion $\kappa(\tau)$. This pair of invariants determine the 
trajectory of a particle up to its position in the space $E_2^1$ [3].

In this part, equations for a component of the metric tensor of the 
world surface of the string at its ends $\dot{x}^2(\tau, \sigma_i)$ 
are derived from boundary equations of a string with masses at ends; 
that tensor defines the torsions of boundary curves and for a 
massless string $(m_i=0)$ it equals zero $\dot x^2(\tau,\sigma_i)=0$. 
It is shown that these nonlinear  equations of second order, when 
$\dot{x}^2(\tau, \sigma_i)$ are periodic, possess constants of motion 
that in some cases allow us to reduce the problem of solution to 
elliptic equations and thus to express $\dot{x}^2(\tau, \sigma_i)$ 
through elliptic functions in a rational way. In the simplest case of 
constant $\dot x^2(\tau,\sigma_i)=c_i$, we arrive at the well--known 
motion of string ends with masses along helixes and the corresponding 
world surface of the string turns out to be a helicoid [4]. In the 
second part, using some examples with different periods $l$ and $2l$,  
we will show how the obtained constants of motion allow us to solve 
the problem of finding the string coordinates.

\section{Equations of motion and boundary conditions}
\setcounter{equation}{0}
Classical equations of motion and boundary conditions for a system of two
point masses connected by the relativistic string follow from the action
function for that system [1, 5]
\begin{equation}
S=- \gamma \int d \tau \int d \sigma \sqrt{{\dot x\acute x}^{2}-{\dot x}^{2}
{\acute x}^{2}} \,
- \, \sum_{i=1}^2 m_i\int d \tau
\sqrt{{\left(
\frac{d x^{\mu}({\tau}_i,{\sigma}_i(\tau))}
                                           { d\tau}\right)}^{2}}
\label{a1}
\end{equation}

Here the first term is the action of a massless relativistic string; 
$\gamma$ is the parameter of tension of the string; $m_i$ are masses 
of particles at the string ends; $x^{\mu}(\tau, \sigma)$ are 
coordinates of the string points in a $D$--dimensional Minkowski 
space with metric $(1, -1, -1, ...)$; derivatives are denoted by

$$
{\dot x}^{\mu}=\frac{\partial x^{\mu}(\tau,\sigma)}{\partial \tau},\,\,
{\acute x}^{\mu}=\frac{\partial x^{\mu}(\tau,\sigma)}{\partial \sigma}
$$

$$
\frac{d x^{\mu}(\tau,\sigma_i(\tau))}{d \tau}=
\dot x^{\mu}(\tau,\sigma_i(\tau))+\acute x^{\mu}(\tau,\sigma_i(\tau))
\dot \sigma_i(\tau),$$
where the string endpoints with masses in the plane of parameters $\tau$ and
$\sigma$ are described by functions $\sigma_i(\tau)$.

As in the case of a massless string, $m_i = 0$, the action (\ref{a1}) is
invariant with respect to a nondegenerate change of parameters
$\tilde \tau=\tilde \tau(\tau,\sigma)$ and
$\tilde \sigma=\tilde \sigma(\tau,\sigma)$, which allows us to take the
conformally flat metric on the string surface by imposing the conditions
of orthonormal gauge
\begin{equation}
\dot x^2+\acute x^2=0,\;\;\;\dot x \acute x=0
\label{a2}
\end{equation}

The action (\ref{a1})
results in the linear equations of motion for the string coordinates [1,2]
\begin{equation}
\ddot x^{\mu}(\tau,\sigma)-x''^{\mu}(\tau,\sigma)=0
\label{a3}
\end{equation}
and the boundary conditions for the ends with masses
\begin{equation}
m_i \frac{d}{d \tau}
\left[ \frac{\dot x^{\mu}(\tau,\sigma_i(\tau))+
\dot \sigma(\tau)\acute x^{\mu}(\tau,\sigma_i(\tau))}
{\sqrt{\dot x^2(\tau,\sigma_i(\tau))(1-{\sigma_i}^2(\tau))}}\right]=
(-1)^{i+1}\gamma
\left[ \dot x^{\mu}(\tau,\sigma_i(\tau))+
\dot \sigma(\tau)\acute x^{\mu}(\tau,\sigma_i(\tau))\right],
\label{a4}
\end{equation}
$$i=1,2.$$

The general solution to equations of motion (\ref{a3}) is the
vector--function
$$x^{\mu}(\tau,\sigma)=1/2\left[\psi^{\mu}_+(\tau+\sigma)+
\psi^{\mu}_-(\tau-\sigma)\right] $$
\begin{equation}
\dot x^{\mu}=1/2[\acute \psi^{\mu}_+(\tau+\sigma)+\acute \psi^{\mu}_-(\tau-
\sigma)];\;\;\;\;
\dot x^{\mu}=1/2[\acute \psi^{\mu}_+(\tau+\sigma)-\acute \psi^{\mu}_-(\tau-
\sigma)],
\label{a5}
\end{equation}
where $\acute \psi^{\mu}_{\pm}(\tau\pm\sigma)$ are derivatives with respect to
the arguments.

Inserting it into the gauge conditions (\ref{a2}) we obtain the equations
\begin{equation}
\acute \psi^2_+(\tau+\sigma)=0,\;\;\;\;\;\acute \psi^2_-(\tau-\sigma)=0,
\label{a6}
\end{equation}
according to which the vectors
$\acute \psi^{\mu}_+(\tau+\sigma) $
and
$\acute \psi^{\mu}_-(\tau-\sigma) $
should be isotropic.  For further consideration, it is convenient to represent
them as expansions over a constant basis in the $D$--dimensional Minkowski
space $E^1_{D_1}$ consisting of two isotropic vectors $a^{\mu}$ and
$c^{\mu}\,\, (a^{\mu}a_{\mu}=0,c^{\mu}c_{\mu}=0,a^{\mu}c_{\mu}=1)$
and $D - 2$ orthonormal space--like vectors $b^{\mu}_k\,\,(r=1,2,3...D-2),
b^{\mu}_k b_{l{\mu}}=-\delta_{kl}$
orthogonal to vectors $a^{\mu}$ and $c^{\mu}\,\,(a^{\mu}b_{k{\mu}}=0,
c^{\mu}b_{k{\mu}}=0)$ [2,7]. As
a result, we obtain the expansion of $\acute \psi^{\mu}_{\pm}$ over this
basis
$$ \acute \psi^{\mu}_+(\tau+\sigma)=\frac{A_+(\tau+\sigma)}
{\sqrt{\sum_{k=1}^{D-2} \dot f^2_k(\tau+\sigma)}}
\left[ a^{\mu}+\sum_{k=1}^{D-2}b^{\mu}_k f_k(\tau+\sigma)+
\frac{1}{2}c^{\mu}\sum_{k=1}^{D-2} f^2_k(\tau+\sigma)\right]$$
\begin{equation}
{}
\label{a7}
\end{equation}
$$
\acute \psi^{\mu}_-(\tau-\sigma)=\frac{A_-(\tau-\sigma)}
{\sqrt{\sum_{k=1}^{D-2} \dot g^2_k(\tau-\sigma)}}
\left[ a^{\mu}+\sum_{k=1}^{D-2}b^{\mu}_k g_k(\tau-\sigma)+
\frac{1}{2}c^{\mu}\sum_{k=1}^{D-2} g^2_k(\tau-\sigma)\right]
$$
It can easily be verified that
$\acute \psi^2_{\pm}=0$,
the condition (\ref{a6}) is satisfied and that
$$\left(\psi''^{\mu}_{\pm}
\psi''_{\pm \mu}\right)=\psi''^2_{\pm}(\tau\pm\sigma)=-A^2_{\pm}(\tau\pm \sigma)$$ where
$A^2_{\pm}(\tau\pm\sigma)$ are two arbitrary functions, like the functions
$f_k$ and $g_k$.  The condition of orthogonal gauge (\ref{a7}) does not
determine the functions $A_{\pm}$, and consequently, there is a possibility to
fix them by imposing further gauge conditions since expressions (\ref{a7}) are
invariant under conformal transformations of the parameters $\tilde \tau \pm
\tilde \sigma=V_{\pm} (\tau \pm \sigma)$.  We fix them by imposing two more
gauge conditions
\begin{equation}
[\ddot x^{\mu}(\tau,\sigma)\pm \dot x'^{\mu}(\tau,\sigma)]^2=-A^2=const,
\label{a8}
\end{equation}
which in terms of the vector--functions
$\acute \psi^{\mu}_{\pm}$ mean
that the space--like vectors $\psi''^{\mu}_{\pm}(\tau\pm\sigma)$ are 
modulo constant,
$$
\psi''^2_{\pm}(\tau+\sigma)=-A^2_{\pm}(\tau\pm\sigma)=-A^2,$$
$$
\phantom{-------------}
(1.8a)
$$

In this way,
we have fixed the functions $A_{\pm}(\tau-\sigma)$ now equal to the constant
$A$.  At the same time, this condition fixes the values of functions
$\sigma_i(\tau)$ (see ref.[2] where it
is shown that $\sigma_i(\tau)=\sigma_i=const $), therefore, we choose
$\sigma_1(\tau)=0$ and $\sigma_2(\tau)=l $.

Further, we will consider the dynamics of a string with masses at the ends
on the plane $(x, y)$, i.e. in the Minkowski space with $D = 3$. In this case,
the expansion (\ref{a7}) contains only one 
space--like vector $b^{\mu}$, and the
expression (\ref{a7}) takes the form
$$\acute \psi^{\mu}_+(\tau+\sigma)=\frac{A}{\acute f(\tau+\sigma)}
[a^{\mu}+b^{\mu}f(\tau+\sigma)+1/2 c^{\mu}f^2(\tau+\sigma)]$$
\begin{equation}
\label{a9}
\end{equation}
$$\acute \psi^{\mu}_-(\tau-\sigma)=\frac{A}{\acute g(\tau-\sigma)}
[a^{\mu}+b^{\mu}f(\tau-\sigma)+1/2 c^{\mu}g^2(\tau-\sigma)],
$$
where $\dot f(\tau+\sigma),\dot g(\tau-\sigma)$ are derivatives with respect to
arguments.
\section{Formulation of boundary equations in terms of  invariants 
of boundary curves}
\setcounter{equation}{0}
Boundary equations (1.4), when $\sigma_i(\tau) = const.$, $\dot x^{\mu}(\tau,
\sigma_i)$ and $\acute x^{\mu}(\tau,\sigma_i).$
>from (\ref{a5}) are substituted into them, and the representation (\ref{a9})
is taken into account, transform into two nonlinear equations for the
functions $f$ and $g$ [2]
$$\frac{d}{d\tau}\ln \left[\frac{\acute g(\tau)}{\acute f(\tau)}\right]
+2 \frac{\acute f(\tau)+\acute g(\tau)}{f(\tau)-g(\tau)}=
\frac{\gamma}{m_1}\left|A\right|\frac{\left |f(\tau)-g(\tau)\right |}
{\sqrt{\acute f(\tau) \acute g(\tau)}},$$
\begin{equation}
{}
\label{b1}
\end{equation}
$$
\frac{d}{d\tau}\ln \left[\frac{\acute g(\tau-l)}{\acute f(\tau+l)}\right]
+2 \frac{\acute f(\tau+l)+\acute g(\tau-l)}{f(\tau+l)-g(\tau-l)}=
-\frac{\gamma}{m_2}\left|A\right|\frac{\left |f(\tau+l)-g(\tau-l)\right |}
{\sqrt{\acute f(\tau+l) \acute g(\tau-l)}}.
$$
whereas nonzero components of the metric tensor of the string surface
$\dot x^2(\tau,\sigma)=-\acute x^2(\tau,\sigma)$
are expressed via $f$ and $g$ as follows [2]
\begin{equation}
\dot x^2(\tau,\sigma)=A^2 \frac{[f(\tau+\sigma)-g(\tau-\sigma]^2}
{4 \acute f(\tau+\sigma)\acute g(\tau-\sigma)}
\label{b2}
\end{equation}
As it is known [1], expression (\ref{b2}) is the general 
solution to the Liouville
equation
$$\frac{\partial^2 ln(\dot x^2(\tau,\sigma))}{\partial^2 \tau}-
\frac{\partial^2 ln(\dot x^2(\tau,\sigma))}{\partial^2 \sigma}=
\frac{A^2}{\dot x^2(\tau,\sigma)},
$$
which in our case is the Gauss equation for the component of metric tensor
of the string minimal surface in the 3--dimensional Minkowski space $E_2^1$
in the gauge (\ref{a2}) and (\ref{a8}). Geometrically [6,7], the conditions
(1.8) mean that the isothermal coordinates (1.2) are at the same time the
asymptotic lines on the string world surface.

>From (\ref{b2}) we obtain the boundary values for the component of metric
tensor $\dot x^2(\tau,\sigma_i)\,\,(\sigma_i=0,l)$
\begin{equation}
\dot x^2(\tau,0)=A^2 \frac{[f(\tau)-g(\tau)]^2}{4 \acute f(\tau)
\acute g(\tau)},\;\;\;\;\;\;
\dot x^2(\tau,l)=A^2 \frac{[f(\tau+l)-g(\tau-l)]^2}{4 \acute f(\tau+l)
\acute g(\tau-l)}.
\label{b3}
\end{equation}
Now we calculate
the curvature $K_i(\tau)$ and torsion $\kappa_i(\tau)$ of boundary curves
along which masses $m_i (i = 1, 2)$ are moving. To this end we compare
boundary equations (\ref{a4}) for $x^{\mu}(\tau,\sigma_i$ in the accepted
gauge
\begin{equation}
\frac{d}{d \tau}\left(\frac{\dot x^{\mu}_i(\tau)}{\sqrt{\dot x^2_i(\tau)}}
\right)=(-1)^{i+1}\frac{\gamma}{m_i}\acute x^{\mu}_i(\tau)\,\,\,\,\,(i=1,2)
\label{b4}
\end{equation}
with the Frenet-Serret  equations [6] for these curves
\begin{equation}
\frac{d}{d \tau}\left(\frac{\dot x^{\mu}_i(\tau)}{\sqrt{\dot x^2_i(\tau)}}
\right)=(-1)^{i+1}K_i(\tau)\acute x^{\mu}_i(\tau)\,\,\,\,\,(i=1,2),
\label{b5}
\end{equation}
\begin{equation}
\frac{d}{d \tau}n^{\mu}_i(\tau)=\kappa_i(\tau)\acute x^{\mu}_i(\tau),
\label{b6}
\end{equation}
where
$x^{\mu}_i(\tau)=x^\mu(\tau,\sigma_i)$, and
$n^{\mu}_i(\tau)=n^{\mu}(\tau,\sigma_i)$  is a unit space--like vector
 of the normal that in the chosen basis
$a^{\mu},b^{\mu},c^{\mu}$ is of the form [2]
$$n^{\mu}(\tau,\sigma)=\frac{2
a^{\mu}+b^{\mu}[f(\tau+\sigma)+g(\tau-\sigma)]+
c^{\mu}f(\tau+\sigma)g(\tau-\sigma)}{f(\tau+\sigma)-g(\tau-\sigma)}.$$
By comparing (\ref{b4}) with (\ref{b5}) we find that the curvatures
$K_i(\tau)$ are constant and equal to
\begin{equation}
K_i(\tau)=\gamma/m_i.
\label{b7}
\end{equation}
Next, projecting (\ref{b6}) onto the vector
$\acute
x^{\mu}_i(\tau)$ and considering that $n^{\mu} (\tau, \sigma)$
is orthogonal to the vectors
$\dot x^{\mu}(\tau,\sigma)$ and $\acute x^{\mu}(\tau,\sigma)$
and $n^2 (\tau, \sigma) = -1$, we obtain
\begin{equation}
\kappa_i(\tau)=\frac{(\dot n_i \acute x_i)}{\acute x^2_i(\tau)}=
\frac{(n_i \dot x'_i)}{\dot x^2_i(\tau)}=\frac{A}{\dot x^2_i(\tau)},\,\,
(i=1,3).
\label{b8}
\end{equation}
Thus, torsions are determined by $\dot x^2_i(\tau)=\dot x^2(\tau,\sigma_i)$
and the constant $A$ that is geometrically a nonzero coefficient of the second
quadratic form outside the string 2--dimensional surface. Indeed, by
definition [6,7]
$$b_{kl}=\left(n\frac{\partial^2 x}{\partial u_k \partial u_l}\right),\quad
\mbox{where}\quad u_1=\tau,\  u_2=\sigma$$,
$$
b_{00}=b_{11}=\frac{1}{2}[ A_+(\tau+\sigma)-A_-(\tau-\tau)],\;\;\;\;\;
b_{01}=b_{10}=\frac{1}{2}[A_+(\tau+\sigma)+A_-(\tau-\tau)].
$$
Therefore, in
our gauge we have $b_{11}=b_{22}=0 $ and $ b_{12}=b_{21}=A$.

Let us turn now to the boundary equations for functions $f$ and $g$ (\ref{b1})
which, in terms of (\ref{b3}), allow us to express in the differential form of
those functions in terms of
the constants $A, K_i$ and the component of
metric tensor $\dot x^2(\tau,\sigma_i)$ on the boundary curves of the string.
For this purpose, we write the r.h.s. of equations (\ref{b1}), in view of
(\ref{b3}), in terms of $K_i$ and $\dot x^2(\tau,\sigma_i)$ as follows
$$
\frac{\gamma}{m_1}A\frac{| f(\tau)-g (\tau)|}{\sqrt{\acute f(\tau)
\acute g(\tau)}}=2 K_1 \sqrt{\dot x^2(\tau,0)}
$$
\begin{equation}
{}
\label{b9}
\end{equation}
$$
\frac{\gamma}{m_2}A\frac{| f(\tau+l)-g (\tau-l)|}{\sqrt{\acute f(\tau+l)
\acute g(\tau-l)}}=2K_2 \sqrt{\dot x^2(\tau,l)}
$$
Then from the
first expression of (\ref{b9}) we express the difference $f(\tau)-g(\tau)$
via the derivatives $\acute f(\tau),\acute g(\tau)$ and $\dot x^2(\tau,0)$;
whereas from the second, the difference $f(\tau+l)-g(\tau-l)$ through the
derivatives $\acute f(\tau+l),\acute g(\tau-l)$ and $\dot x^2(\tau,l)$:
$$
f(\tau)-g(\tau)=\epsilon[f(\tau)-g(\tau)]\frac{1}{A}\sqrt{\acute f(\tau)
\acute g(\tau)\dot x^2(\tau,0)},
$$
\begin{equation}
{}
\label{b10}
\end{equation}
$$
f(\tau+l)-g(\tau-l)=\epsilon[f(\tau+l)-g(\tau-l)]\frac{1}{A}\sqrt{\acute
f(\tau+l) \acute g(\tau-l)\dot x^2(\tau,l)},
$$
where $\epsilon[x]$ is the signum function:
$$\epsilon[x] = \left\{ \begin{array}{rcl}
-1, x<0\\
1, x>0\\
\end{array}\right..
 $$  From
(\ref{b3}) it follows that in view of $\dot x^2(\tau,\sigma)>0  $
when $m_i \neq 0$, then $\acute f(\tau,\sigma) \acute
g(\tau,\sigma)>0$ throughout.  Eliminating the difference $f - g$
>from (\ref{b1}) with the use of (\ref{b10}), we obtain the boundary
equations containing only the derivatives of functions $\acute
f,\acute g)$ and $\sqrt {\dot
x^2(\tau,\sigma_i)}$:
$$
\frac{d}{d \tau}ln\left[\frac{\acute
g(\tau)}{\acute f(\tau)}\right]+ \frac{A\epsilon_1}{\sqrt{\dot
x^2(\tau,0)}} \left(\sqrt{\frac{\acute f(\tau)}{\acute
g(\tau)}}+\sqrt{\frac{\acute g(\tau)}{\acute f(\tau)}}\right)=2
K_1\sqrt{\dot x^2(\tau,0)},\phantom{---------}(2.11a)
$$
$$
\frac{d}{d
\tau}ln\left[\frac{\acute g(\tau-l)}{\acute f(\tau+l)}\right]+
\frac{A\epsilon_2}{\sqrt{\dot x^2(\tau,l)}} \left(\sqrt{\frac{\acute
f(\tau+l)}{\acute g(\tau-l)}}+\sqrt{\frac{\acute g(\tau-l)}{\acute
f(\tau+l)}}\right)=-2 K_2\sqrt{\dot x^2(\tau,l)},\phantom{---}(2.11b)
$$
where $\epsilon_1=\epsilon[\acute
f(\tau)\{f(\tau)-g(\tau)\}], \epsilon_2=\epsilon[\acute
f(\tau+l)\{f(\tau+l)-g(\tau-l)\}].$

Together with this system of
boundary equations, we also consider equalities arising upon the
calculation of the logarithmic derivative of (\ref{b3}); in this way,
with (\ref{b10}), we get
$$
\frac{d}{d \tau}ln\left[\acute
g(\tau)\acute f(\tau)\right]- \frac{A\epsilon_1}{\sqrt{\dot
x^2(\tau,0)}} \left(\sqrt{\frac{\acute f(\tau)}{\acute
g(\tau)}}-\sqrt{\frac{\acute g(\tau)}{\acute
f(\tau)}}\right)=-\frac{d}{d \tau}\ln\dot x^2(\tau,0),\phantom{-------}(2.12a)
$$
$$
\frac{d}{d \tau}ln\left[\acute g(\tau-l)\acute
f(\tau+l)\right]- \frac{A\epsilon_2}{\sqrt{\dot x^2(\tau,l)}}
\left(\sqrt{\frac{\acute f(\tau+l)}{\acute
g(\tau-l)}}-\sqrt{\frac{\acute g(\tau-l)}{\acute
f(\tau+l)}}\right)=-\frac{d}{d \tau}\ln\dot x^2(\tau,l).\phantom{--}(2.12b)
$$
\addtocounter{equation}{2}

The sum and difference of the equations (2.11a) and (2.11b) give two
equations
$$
\left[2\frac{d}{d \tau}+K_1 \sqrt{\dot x^2(\tau,0)}-\frac{d}{d
\tau}\ln \sqrt{\dot x^2(\tau,0)}\right]\frac{1}{\sqrt{\acute
g(\tau)}}= \frac{A\epsilon_1}{\sqrt{\dot x^2(\tau,0)\acute f(\tau)}},
$$
\begin{equation}
{}
\label{b13}
\end{equation}
$$
\left[2\frac{d}{d \tau}-K_1 \sqrt{\dot
x^2(\tau,0)}-\frac{d}{d \tau}\ln \sqrt{\dot
x^2(\tau,0)}\right]\frac{1}{\sqrt{\acute f(\tau)}}=
-\frac{A\epsilon_1}{\sqrt{\dot x^2(\tau,0)\acute g(\tau)}}
$$
for the first boundary ($\sigma_1=0$). In
a similar way, the sum and difference of the equations (2.11b)
 and (2.12b) of the
same systems provide two equations
$$
\left[2\frac{d}{d \tau}-K_2
\sqrt{\dot x^2(\tau,l)}-\frac{d}{d \tau}\ln \sqrt{\dot
x^2(\tau,l)}\right]\frac{1}{\sqrt{\acute g(\tau-l)}}=
\frac{A\epsilon_2}{\sqrt{\dot x^2(\tau,l)\acute f(\tau+l)}},
$$
\begin{equation}
\left[2\frac{d}{d \tau}+K_2 \sqrt{\dot
x^2(\tau,l)}-\frac{d}{d \tau}\ln \sqrt{\dot
x^2(\tau,l)}\right]\frac{1}{\sqrt{\acute f(\tau+l)}}=
-\frac{A\epsilon_1}{\sqrt{\dot x^2(\tau,l)\acute g(\tau-l)}}
\label{b14}
\end{equation}
for the second boundary
($\sigma_2=l$).

Now let us  derive equations separately for the functions
$f(\tau)$ and  $g(\tau)$ and, respectively, for $f(\tau+l)$ and $g(\tau-l)$.
This is achieved by eliminating $1/\sqrt{g'(\tau)}$ from (\ref{b13});
thus, for $1/\sqrt{f'(\tau)}$ we obtain
\begin{equation}
D\left[f(\tau)\right]=D\left[A\int_0^{\tau}\frac{d \varsigma}{\sqrt{\dot
x^2(\varsigma,0)}}\right]+\frac{A K_1}{2}\left(\frac{A}{K_1 \dot x^2(\tau,0)}
-\frac{K_1 \dot x^2(\tau,0)}{A}\right)-2 K_1 A\frac{d}{d \tau}
\sqrt{\dot x^2(\tau,0)},
\label{b15}
\end{equation}
where $D[y(x)]$ is the Schwartz derivative defined by
\begin{equation}
D[y(x)]=-2\sqrt {\acute y(x)}\frac{d^2}{d x^2}\left(\frac{1}
{\sqrt {\acute y(x)}}\right)=\frac{y'''(x)}{y'(x)}-
\frac{3}{2}\left(\frac{y''(x)}{y'(x)}\right)^2.
\label{b16}
\end{equation}
Then, removing $1/\sqrt{f'(\tau)}$ from (\ref{b13}) we
arrive at the equation for $1/\sqrt{g'(\tau)}$
\begin{equation}
D\left[g(\tau)\right]=D\left[A\int_0^{\tau}\frac{d \eta}{\sqrt{\dot
x^2(\eta,0)}}\right]+\frac{A K_1}{2}\left(\frac{A}{K_1 \dot x^2(\tau,0)}
-\frac{K_1 \dot x^2(\tau,0)}{A}\right)+2 K_1 A\frac{d}{d \tau}
\sqrt{\dot x^2(\tau,0)}.
\label{b17}
\end{equation}
The same procedure
for system (\ref{b14}) results in the equations for $1/\sqrt{f'(\tau+l)}$
and \\ $1/\sqrt{g'(\tau-l)}$ for the second boundary ($\sigma_2=l)$:
\begin{equation}
D\left[f(\tau+l)\right]=D\left[A\int_{}^{\tau}\frac{d \eta}{\sqrt{\dot
x^2(\eta,l)}}\right]+\frac{A K_2}{2}\left(\frac{A}{K_2 \dot x^2(\tau,0)}
-\frac{K_2 \dot x^2(\tau,l)}{A}\right)+2 K_2 A\frac{d}{d \tau}
\sqrt{\dot x^2(\tau,l)},
\label{b18}
\end{equation}
\begin{equation}
D\left[g(\tau-l)\right]=D\left[A\int_{}^{\tau}\frac{d \eta}{\sqrt{\dot
x^2(\eta,l)}}\right]+\frac{A K_2}{2}\left(\frac{A}{K_2 \dot x^2(\tau,l)}
-\frac{K_2 \dot x^2(\tau,l)}{A}\right)-2 K_2 A\frac{d}{d \tau}
\sqrt{\dot x^2(\tau,l)}.
\label{b19}
\end{equation}
\bigskip

 From
these equations it follows that the functions $f(\tau)$and$g(\tau)$
are defined
by $\dot x^2(\tau,0)$ in accordance with (\ref{b15}) and (\ref{b17}); whereas
$f(\tau+l)$ and $g(\tau-l)$, by $\dot x^2(\tau,l)$ according to (\ref{b18})
and (\ref{b19}) since the r.h.s.  of these equations contain only
$\dot x^2(\tau,\sigma_i)$  and constants $A, K_i$.

For $\dot x^2(\tau,0)$ and $\dot x^2(\tau,l)$ we can obtain equations that
connect them with each other by changing the argument $\tau$ in eq.
(\ref{b18}) to $\tau - l$; and in
(\ref{b19}) to $\tau + l$, we find that the left--hand sides
of eqs.  (\ref{b15}) and (\ref{b18}), as well as (\ref{b17}) and
(\ref{b19}) coincide.  As a result, we arrive at the two equations
$$
D\left[f(\tau)\right]=D\left[A\int_{}^{\tau}\frac{d \eta}{\sqrt{\dot
x^2(\eta,0)}}\right]+\frac{A K_1}{2}\left(\frac{A}{K_1 \dot x^2(\tau,0)}
-\frac{K_1 \dot x^2(\tau,0)}{A}\right)-2 K_1 A\frac{d}{d \tau}
\sqrt{\dot x^2(\tau,0)}=
$$
\begin{equation}
D\left[A\int_{}^{\tau-l}\frac{d \eta}{\sqrt{\dot
x^2(\eta,l)}}\right]+\frac{A K_2}{2}\left(\frac{A}{K_2 \dot x^2(\tau-l,l)}
-\frac{K_2 \dot x^2(\tau-l,l)}{A}\right)+2 K_2 A\frac{d}{d \tau}
\sqrt{\dot x^2(\tau-l,l)},
\label{b20}
\end{equation}
\bigskip
$$
D\left[g(\tau)\right]=D\left[A\int_{}^{\tau}\frac{d \eta}{\sqrt{\dot
x^2(\eta,0)}}\right]+\frac{A K_1}{2}\left(\frac{A}{K_1 \dot x^2(\tau,0)}
-\frac{K_1 \dot x^2(\tau,0)}{A}\right)+2 K_1 A\frac{d}{d \tau}
\sqrt{\dot x^2(\tau,0)}=
$$
\begin{equation}
D\left[A\int_{}^{\tau+l}\frac{d \eta}{\sqrt{\dot
x^2(\eta,l)}}\right]+\frac{A K_2}{2}\left(\frac{A}{K_2 \dot x^2(\tau+l,l)}
-\frac{K_2 \dot x^2(\tau+l,l)}{A}\right)-2 K_2 A\frac{d}{d \tau}
\sqrt{\dot x^2(\tau+l,l)}.
\label{b21}
\end{equation}

The
second equalities in (\ref{b20}) and (\ref{b21}) represent just the
connection
between $\dot x^2(\tau,0)$ and $\dot x^2(\tau,l)$.

Further, from (\ref{b15}) and (\ref{b17}) it follows that the difference of
the Schwartz derivatives of the functions $f(\tau)$ and $g(\tau)$ is given by
\begin{equation}
D[f(\tau)]-D[g(\tau)]=-4 A K_1 \frac{d}{d \tau}\sqrt{\dot x^2(\tau,0)},
\label{b22}
\end{equation}
and from(\ref{b18})and (\ref{b19})
\begin{equation}
D[f(\tau+l)]-D[g(\tau-l)]=4 A K_2 \frac{d}{d \tau}\sqrt{\dot x^2(\tau,l)},
\label{b23}
\end{equation}
Eliminating $D[g(\tau)]$ from these equations by the change of $\tau$ to
$\tau +l$ in the  equation (\ref{b23}) and then eliminating $D[f(\tau)]$ by the
change of
$\tau$ to $\tau - l$ in the equation (\ref{b23}), we obtain the equations
$$
D[f(\tau+2l)]-D[f(\tau)]=4 A \frac{d}{d \tau}[K_1\sqrt{\dot x^2(\tau,0)}+
K_2\sqrt{\dot x^2(\tau+l,l)}],
$$
\begin{equation}
\label{b24}
\end{equation}
$$
D[g(\tau)]-D[g(\tau-2l)]=4 A \frac{d}{d \tau}[K_1\sqrt{\dot x^2(\tau,0)}+
K_2\sqrt{\dot x^2(\tau-l,l)}],
$$
whose left--hand sides
contain either the function $f$ or $g$ with shifted arguments, whereas the
right--hand sides depend on $\sqrt{\dot x^2(\tau,0)}$ and $\sqrt{\dot
x^2(\tau\pm l,l)}$.  These equations
give conserved quantities when the difference of Schwarz derivatives on the
 left--hand sides are zero under certain
conditions of periodicity to be considered in sect 3.

A simplest example of the solution of boundary equations within this approach
is the case of constant $\dot x^2(\tau,\sigma_i)=\dot x^2_{0i}$, i.e. constant
torsions of boundary curves $\kappa(\tau,\sigma_i)=\kappa_{0i}$ according to
(\ref {b8})
  It is known [3],[6] that curves in a three--dimensional space with a
constant curvature $K_i$ and a constant torsion $\kappa_{0i}$ are helixes, and
the minimal surface within these boundaries is a helicoid. A detailed
solution of the corresponding boundary equations is presented in author's
paper [4]; here we briefly outline the way of solution of that problem in the
given approach.  From equations (\ref{b22})--(\ref{b24}) at $\dot
x^2_{0i}=const$ right-hand sides of these equations are zero and
 we obtain the equalities
$$
D[f(\tau)]=D[g(\tau)];\qquad
D[f(\tau+l)]=D[g(\tau-l)];
$$
$$D[f(\tau+2l)]=D[f(\tau)];\qquad
D[g(\tau)]=D[g(\tau+2l)],
$$
>from which it follows
(see Appendix A) that the functions entering into the Schwartz derivatives are
connected with each other by linear--fractional expressions; in particular,
>from the first and second equalities it follows that
\begin{equation}
g(\tau)=\frac{a_1 f(\tau)+b_1}{c_1 f(\tau)+d_1},\quad
g(\tau-l)=\frac{a_2 f(\tau+l)+b_2}{c_2 f(\tau+l)+d_2},
\label{b25}
\end{equation} where $a_i,b_i,c_i,d_i$ are
arbitrary constants such that $a_i d_i-b_i c_i= 1$.  From equations
(\ref{b20}) and (\ref{b21}) we get
\begin{equation}
D[f(\tau)]=D[g(\tau)]=\frac{A K_1}{2}\left(\frac{A}{K_1 \dot x^2_{01}}-
\frac{K_1 \dot x^2_{01}}{A}\right)=
\frac{A K_2}{2}\left(\frac{A}{K_2 \dot x^2_{02}}-
\frac{K_2 \dot x^2_{02}}{A}\right).
\label{b26}
\end{equation}
Denoting constant quantities equal
to each other by $\omega$ that, according to [4], is the angular velocity of
rotation of a rectilinear string around the center of rotation, we have the
formula
$$
K_i\left(\frac{A}{K_i \dot x^2_{0i}}-\frac{K_i \dot x^2_{0i}}{A}\right)=
\omega
$$ from which we can express $\dot x^2_{0i}$ and the torsion $\kappa_{0i}$ in
terms of $\omega,A,K_i$
\begin{equation}
\kappa_{0i}=\frac{A}{\dot x^2_{0i}}=K_i\left(\sqrt{{\left(\frac{\omega}{2
K_i} \right)}^2+1}+\frac{\omega}{2 K_i}\right)
\label{b27}
\end{equation}
Instead of
equations (\ref{b26}) that are linear equations of the second order in
$1/\sqrt{\acute f(\tau)}$
and
$1/\sqrt{\acute g(\tau)}$
, it is easier to determine the functions $f(\tau)$ and $g(\tau)$ from the
initial boundary equations (\ref{b1}) since they are equations of the second
order in $f(\tau)$ and $g(\tau)$ and the ratios of derivatives
$g'(\tau)/f'(\tau)$ and $g'(\tau-l)/f'(\tau+l)$ are, according to
(\ref{b25}), equal to
$[c_1 f(\tau)+d_1]^{-2}$
and
$[c_2 f(\tau+l)+d_2]^{-2}$
, resp., which reduces
equations (2.1) to two equations of the first order whose solution fixes
the constants $a_i,b_i,c_i,d_i$ in terms of $A,K_i$ and $\omega$.  (In ref.
[4], the energy $E$ and angular momentum $J$ have been calculated for such a
rotating rectilinear string with a given angular velocity $\omega$ and masses
$m_i$ at the ends).

\section{Constants of motion for boundary equations of a string with periodic
torsions of trajectories of ends }
\setcounter{equation}{0}

It is a remarkable fact that the system of boundary equations (\ref{b20}) and
(\ref{b21}) possesses conserved quantities when $\dot x^2(\tau,\sigma_i)$ are
periodic with a period multiple of $l$: \\ $\dot x^2(\tau,\sigma_i)=
\dot x^2(\tau+n l,\sigma_i), n=1,2,3...$; in this
case, the torsions of boundary curves will also be periodic,
$\kappa_i(\tau)=\kappa_i(\tau+n l)$.

The right--hand sides of the above
equations depend only on $\dot x^2(\tau,\sigma)$, consequently, their
left--hand sides should be periodic with the same period:
\begin{equation}
D[f(\tau+n l)]=D[f(\tau)],\quad
D[g(\tau+n l)]=D[g(\tau)].
\label{c1}
\end{equation}
 In view of
the property of the Schwartz derivative (see Appendix A) we have
$$f(\tau+n l)=\frac{a_1 f(\tau)+b_1}{c_1 f(\tau)+d_1}=T_1 f(\tau), $$
\begin{equation}
\label{c2}
\end{equation}
$$
g(\tau+n l)=\frac{a_2 g(\tau)+b_2}{c_2 g(\tau)+d_2}=T_2 g(\tau).
$$
We
will prove that  these two linear--fractional
transformations are to be equal: $T_1=T_2$ . \\
To this end, using (\ref{c2})
and (\ref{b3}), we write the condition of periodicity for $\dot
x^2(\tau,\sigma_i)$
$$
\dot x^2(\tau,0)=\frac{A^2[f(\tau)-g(\tau)]^2}{4 \acute f(\tau)
\acute g(\tau)}=\dot x^2(\tau+n l,0)=\frac{A^2[T_1 f(\tau)-T_2 g(\tau)]^2}
{4 (T_1 f(\tau))'(T_2 f(\tau))'};
$$
\begin{equation}
\label{c3}
\end{equation}
$$
\dot x^2(\tau,l)=\frac{A^2[f(\tau+l)-g(\tau-l)]^2}{4 \acute f(\tau+l)
\acute g(\tau-l)}=\dot x^2(\tau+n l,l)=\frac{A^2[T_1 f(\tau+l)-T_2
g(\tau-l)]^2} {4 (T_1 f(\tau+l))'(T_2 f(\tau-l))'}.
$$
Since the derivatives of the linear--fractional
function are given by the expressions
$$
\left(T_1 f(\tau)\right)'=\frac{f'(\tau)}{[c_1 f(\tau)+d_1]^2},\quad
\left(T_2 g(\tau)\right)'=\frac{g'(\tau)}{[c_2 g(\tau)+d_2]^2},
$$
 the denominators in (\ref{c3})
coincide, and the numerators obey the equality
$$
[f(\tau)-g(\tau)]=(a_1 f(\tau)+b_1)(c_2 g(\tau)+d_2)-(c_1 f(\tau)+d_1)
(a_2 g(\tau)+b_2)
$$ and the same equality
follows from the second eq. of (\ref{c3}) but with shifted arguments of
$f(\tau+l)$ and $g(\tau-l)$. These equalities, provided that $a_i d_i-b_i c_i=
1$, hold valid under the condition $$
a_1=a_2=a,\;b_1=b_2=b,\;c_1=c_2=c,\;d_1=d_1=d.
$$ Thus the periodicity condition (\ref{c3}) results in
that  $f$ and $g$ are transformed as follows
\begin{equation}
f(\tau+n l)=T f(\tau),\quad g(\tau+nl)=T g(\tau),\;\mbox{where}\;
T f(\tau)=\frac{a f(\tau)+b}{c f(\tau)+d}.
\label{c4}
\end{equation}

Now we can consider each of periods $l, 2l, ..., nl$ separately and
consequences  that follow from eqs. (\ref{b24}) in these cases.

For the period $l$, ${\dot x}^2(\tau+l,\sigma_i)={\dot x}^2(\tau,\sigma_i)$
>from eq. (\ref{c4}) it follows that $f(\tau+l)=Tf(\tau)$
and $g(\tau-l)=T^{-1}g(\tau)$, where $T^{-1}$ is the inverse
linear--fractional transformation, and
$$
f(\tau+2l)=T(T f(\tau))=\frac{(a^2+c b)f(\tau)+b(a+d)}{c(a+d)f(\tau)+d^2+c b}=
\frac{\left[a-(a+d)^{-1}\right]f(\tau)+b}{c f(\tau)+d-(a+d)^{-1}},
$$
$$\phantom{--}
g(\tau-2l)=T^{-1}(T^{-1} g(\tau))=\frac{(d^2+c
b)g(\tau)-b(a+d)}{-c(a+d)g(\tau)+a^2+c b}=
\frac{\left[d-(a+d)^{-1}\right]g(\tau)-b} {-c g(\tau)+a-(a+d)^{-1}}\;\;\;\;\;
$$ are also linear--fractional transformations with the determinant equal to
unity when $ad - bc = 1$. Then, the left--hand side of eqs. (\ref{b24}) are
zero because
$$D[f(\tau+2l)]=D[f(\tau)],\quad
D[g(\tau)]=D[g(\tau-2l)],
$$ and from (\ref{b24}) we obtain the conserved quantity for the motion with
$
{\dot x}^2(\tau,\sigma_i)={\dot x}^2(\tau+l,\sigma_i)
$
\begin{equation}
K_1\sqrt{\dot x^2(\tau,0)}+K_2\sqrt{\dot x^2(\tau,l)}=h_1,
\label{c5}
\end{equation}
 where $h_1$ is the constant of integration.

For equal masses at the string ends $m_1=m_2,\;K_1=K_2=\gamma/m$, and if we
put $f(\tau+l)=g(\tau)\;$and$\;g(\tau-l)=f(\tau)$, the equality
$\dot x^2(\tau,0)=\dot x^2(\tau,l)$ is
fulfilled and the second boundary (2.1) turns into the first one. Then
equation (3.5) results in constant $\dot x^2(\tau,\sigma_i)$ because
$$
K\sqrt{\dot x^2(\tau,0)}
=K\sqrt{\dot x^2(\tau,l)}=h_1/2
$$

Now let us consider the case with period $2l$: $\dot x^2(\tau+2l,\sigma_i)=
\dot x^2(\tau,\sigma_i)$.
According to (\ref{c4}), $f(\tau+2l)=T f(\tau)$ and $g(\tau-2l)=T^{-1} g(\tau)$,
therefore,
$D[f(\tau+2l)]=D[f(\tau)]$
and
$D[g(\tau-2l)]=D[g(\tau)]$,
then the left-hand sides of eqs.  (\ref{b24}) again turn
out to be zero; upon integration we obtain
\begin{equation}
K_1\sqrt{\dot x^2(\tau,0)}
+K_2\sqrt{\dot x^2(\tau\pm l,l)}= h_2.
\label{c6}
\end{equation}
This constant of motion for the period $2l$ differs from (3.5) by the argument
in the second term shifted by $l$. Therefore, when masses are equal, $K_1=K_2$,
and the special case, $\dot x^2(\tau,0)=\dot x^2(\tau,l)=\dot x^2(\tau),\;
f(\tau+l)=g(\tau)$, is considered, we do not obtain constant $\dot x^2(\tau,
\sigma_i)$
since in this special case (\ref{c6}) results in the expression
\begin{equation}
K\left[\sqrt{\dot x^2(\tau)}+\sqrt{\dot x^2(\tau\pm l)}\right]=h_2
\label{c7}
\end{equation}
which is fulfilled not only for constant $\dot x^2(\tau)$.
The derivation of $\dot x^2(\tau,\sigma)$  and solution
of the whole problem for the period $2l$ will be done
 in a second part of a  paper (II) .

The case with period $3l$ is more complicated.  For this period, from the
first of eqs.(\ref{b24}), by shifting the argument $\tau$ by $l$ and then by
$2l$, we obtain the system of three equations:
$$
D[f(\tau+2l)]-D[f(\tau)]=4 A \frac{d}{d \tau}\left[K_1 \sqrt{\dot x^2(\tau,0)}
+K_2 \sqrt{\dot x^2(\tau+l,l)}\right];
$$
\begin{equation}
D[f(\tau+3l)]-D[f(\tau+l)]=4 A \frac{d}{d \tau}\left[K_1 \sqrt
{\dot x^2(\tau+l,0)}
+K_2 \sqrt{\dot x^2(\tau+2l,l)}\right];
\label{c8}
\end{equation}
$$
D[f(\tau+4l)]-D[f(\tau+2l)]=4 A \frac{d}{d \tau}\left[K_1 \sqrt
{\dot x^2(\tau+2l,0)}
+K_2 \sqrt{\dot x^2(\tau+3l,l)}\right].
$$
Summing these equalities and considering that $\;D[f(\tau+3l)]=D[f(\tau)],
\;D[f(\tau+4l)]=D[f(\tau+l)]\;$ and$\;\dot x^2(\tau+3l,l)=\dot x^2(\tau,l)
$, we
get
$$
0=4A \frac{d}{d \tau}\left\{
K_1\left(\sqrt{\dot x^2(\tau,0)}+
\sqrt{\dot x^2(\tau+l,0)}+\sqrt{\dot x^2(\tau+2l,0)}\right)+\right.
$$
$$
\qquad {} \qquad {} \left .K_2\left(\sqrt{\dot x^2(\tau,l)}+
\sqrt{\dot x^2(\tau+l,l)}+\sqrt{\dot x^2(\tau+2l,l)}\right)\right\}
$$
and upon integration we have
\begin{equation}
\sum_{m=0}^2\left[K_1 \sqrt{\dot x^2(\tau+ml,0)}+K_2 \sqrt{\dot x^2(\tau+
ml,l)}\right]=h_3
\label{c9}
\end{equation}
In the same way, from the
second of eqs.(\ref{b24}), by shifting the argument $\tau$ by $-l$, and
then by
$-2l$, we obtain three equations the sum of which gives
\begin{equation}
   4A \sum_{m=0}^2\frac{d}{d\tau}\left[K_1 \sqrt{\dot x^2(\tau-ml,0)}+K_2 
\sqrt{\dot x^2(\tau-
ml,l)}\right]=0
\label{c10}
\end{equation}
This
expression coincides with (\ref{c9}) when $\tau$ is changed to $\tau + 2l$.

>From these examples it is not difficult to deduce the general expression for
a conserved quantity for period $nl$ that is different for even and odd $n$.

For even $n = 2r (r = 1, 2,...)$, it is necessary, upon adding $2ml$ to the
argument $\tau$ in eq. (\ref{b24}), to sum up the obtained expressions over
$m$
>from zero to $r - 1$, which gives
$$
     \sum_{m=0}^{r-1}D[f(\tau+2(1+m)l)]-\sum_{m=0}^{r-1}D[f(\tau+2ml)]=
$$
$$
   4A  \sum_{m=0}^{r-1}\frac{d}{d \tau}\left\{K_1 \sqrt{\dot x^2(\tau+2ml,0)}+
K_2 \sqrt{x^2(\tau+(1+2m)l,l)}\right\}.
$$
The left--hand side of this equation equals zero since under the change
$1+m=m'$ in the first sum we have
$$
\sum_{m'=1}^{r}D[f(\tau+2m'l)]-\sum_{m=0}^{r-1}D[f(\tau+2ml)]=
D[f(\tau+2rl)]-D[f(\tau)]=0,
$$
and hence the constant quantity is
\begin{equation}
\sum_{m=0}^{r-1}\left\{K_1 \sqrt{\dot x^2(\tau+2ml,0)}+
K_2 \sqrt{x^2(\tau+l+2ml,l)}\right\}=h_{2r}.
\label{c11}
\end{equation}
When $r = 1$, from (\ref{c11}) we obtain (\ref{c6}) with period $2l$.

For odd $n = 2r + 1 (r = 0, 1, 2, ...)$, it is necessary, adding $ml$ to the
argument in (\ref{b24}), to sum up the equations over $m$ from zero to $2r$,
then
$$
\sum_{m=0}^{2r}D[f(\tau+2l+ml)]-\sum_{m=0}^{2r}D[f(\tau+ml)]=
$$
\begin{equation}
   4A \sum_{m=0}^{2r}\frac{d}{d \tau}\left\{K_1 \sqrt{\dot x^2(\tau+ml,0)}+
K_2 \sqrt{x^2(\tau+l+ml,l)}\right\}.
\label{c12}
\end{equation}

Again, the left--hand side of the equation (3.12) is zero since setting
$2+m=m'$ in the first sum and considering that $(1+2k)l$ is a period, we get
$$
\sum_{m'=2}^{2+2r}D[f(\tau+m'l)]-\sum_{m=0}^{2r}D[f(\tau+ml)]=
$$
$$D[f(\tau+(1+2r)l)]-D[f(\tau+2(1+r)l)]-D[f(\tau)]-D[f(\tau+l)]=0,
$$

Consequently, in this case the quantity
\begin{equation}
\sum_{m=0}^{2r}\left\{K_1 \sqrt{\dot x^2(\tau+ml,0)}+
K_2 \sqrt{x^2(\tau+ml,l)}\right\}=h_{2r+1}.
\label{c13}
\end{equation}
is constant.  In
(\ref{c13}) we considered that the last term in the sum of the second term
in (\ref{c12}) equals $\dot x^2(\tau+l+2rl,l)=\dot x^2(\tau,l)$. When $r = 0$
and $r = 1$, we obtain (\ref{c5}) and (\ref{c6}).

So, (\ref{c11}) and (\ref{c13}) are constants of motion of the boundary
equations of a relativistic string with masses at ends when masses are moving
along the curves with periodic torsion $\kappa_i(\tau+nl)=\kappa_i(\tau)$ and
constant curvature $K_i=\gamma/m_i$.  The curves with constant curvatures in
the Euclidean geometry are called the Bertrand curves [6]; in our case, when
$K_1=K_2(m_1=m_2)$, two boundary curves along which the masses $m_i$ are moving
are two conjugate Bertrand curves, i.e. the center of curvature of one curve
is always on the other curve.

\section{Conclusion}
\setcounter{equation}{0}
The obtained constants of
motion (\ref{c11}) and (\ref{c13}) can geometrically be interpreted as
follows:  Since the length of a curve $L_i$ between points $\tau_2$ and
$\tau_1$ is given by the expression
\begin{equation}
L_i(\tau_2,\tau_1)=\int\limits_{\tau_2}^{\tau_1}
\sqrt{\dot x^2(\tau,\sigma_i)}d \tau,
\label{d1}
\end{equation}
then, integrating (\ref{c11}), (\ref{c13})
in the interval [$\tau_2, \tau_1$] and expressing the curvature $K_i$ through
the curvature radius $R_i$: $ R_i=1/K_i$, we get
\begin{equation}
\sum_{m=0}^{r-1} \left[\frac{1}{R_1}L_1(\tau_1+2ml,\tau_2+2ml)+\frac{1}{R_2}
L_2(\tau_1+2ml,\tau_2+2ml)\right]=h_{2r}(\tau_1-\tau_2),
\label{d2}
\end{equation}
\begin{equation}
\sum_{m=0}^{2r}\left[\frac{1}{R_1}L_1(\tau_1+ml,\tau_2+ml)+\frac{1}{R_2}
L_2(\tau_1+ml,\tau_2+ml)\right]=h_{2r+1}(\tau_1-\tau_2),
\label{d3}
\end{equation}
>From these
expressions it is seen that sums of the curves divided by constant
 radii $R_i$ of
their curvatures grow linearly with the parameter $\tau$ as though their
element of length  were a constant $\sqrt{\dot x^2_{0i}}$.
Consequently, we can set the constant $h_{2r}$ in (\ref{c11}) to be equal to
$$
h_{2r}=r\left(K_1\sqrt{\dot x^2_{01}}+K_2\sqrt{\dot x^2_{02}}\right),
$$
whereas the constant $h_{2r+1}$ in (\ref{c13}), to
$$
h_{2r+1}=(2r+1)\left(K_1\sqrt{\dot x^2_{01}}+K_2\sqrt{\dot x^2_{02}}\right).
$$

In particular,
for the period $l$ in (3.5) $r = 0$, and for the period $2l$ in (3.6) $r = 1$,
therefore,
$$
h_{1}=h_2=K_1\sqrt{\dot x^2_{01}}+K_2\sqrt{\dot x^2_{02}}.
$$

It has been mentioned that the constants $\sqrt{\dot x^2(\tau,\sigma_i)}$
arise only when masses $m_i$ move along helixes and thus the sum of lengths
along which $m_i$ pass in the case of periodic motion during the intervals of
$\tau$ multiple of $l$ up to $nl$ is equal to the length passed by the same
point $m_i$ as though it was moving along a helix with constant $\dot
x^2_{01}$.

In part II, we will show how the integrals of motion we have here derived
can be applied to find the world surface of a string when $\dot
x^2(\tau,\sigma_i)$ has periods $l$ and $2l$. These solutions are expressed
through elliptic functions and describe motion of the relativistic string that
is more complicated than rotation of the string as a finite straight line,
therefore, the string world surfaces for both periods $l$ and $2l$ do not
belong to the class of ruled surfaces. A solution of that sort describes
transverse excitations of the string and radial motions of masses.
\begin{center}
 {\bf Acknowledgement.}
\end{center}
The author expresses his gratitude to
V.V.Nesterenko and A.M.Chervyakov for useful discussions. He is also
grateful to A.L.Koshkarov and I.G.Pirozhenko for help in preparing this
material for publication. The work
is supported by the Russian Foundation for Fundamental Research, grant No
96--02--00556.

\section*{Appendix A}
\setcounter{equation}{0}
An important property of the Schwartz derivative $D[f(\tau)]$ defined by
$$
D[f(\tau)]=\frac{f'''(\tau)}{f'(\tau)}-\frac{3}{2}\left(\frac{f''(\tau)}
{f'(\tau)}\right)^2=-2
\sqrt{f'(\tau)}\frac{d^2}{d
\tau^2}\left(\frac{1}{\sqrt{f'(\tau)}}\right),\phantom{-------}(A.1)
$$
is that it is
invariant under linear--fractional transformations of the function $f(\tau)$
$$
f(\tau)\to\frac{a f(\tau)+b}{c f(\tau)+d},\quad  ad-bc=1,
$$ which is easily proved by using the second form for $D[f(\tau)]$ given in
(A.1) and considering that $f'(\tau)\to f'(\tau)[c f(\tau)+d]^{-2}$.

One more
important property of these derivatives consists in that from the equality of
the Schwartz derivatives of two functions $f(\tau)$ and $g(\tau)$ it follows
that these functions are connected with each other via a linear--fractional
transformation. Indeed, from $D[f(\tau)]=D[g(\tau)]$ it follows that
$$
\sqrt{f'(\tau)}\frac{d^2}{d
\tau^2}\left(\frac{1}{\sqrt{f'(\tau)}}\right)=
\sqrt{g'(\tau)}\frac{d^2}{d
\tau^2}\left(\frac{1}{\sqrt{g'(\tau)}}\right)
$$
or, assuming that $f'(\tau),g'(\tau)\ne0$, we have
$$
0=\frac{1}{\sqrt{g'(\tau)}}\frac{d^2}{d
\tau^2}\left(\frac{1}{\sqrt{f'(\tau)}}\right)-
\frac{1}{\sqrt{f'(\tau)}}\frac{d^2}{d
\tau^2}\left(\frac{1}{\sqrt{g'(\tau)}}\right)=
$$
$$
\frac{d}{d \tau}\left[\frac{1}{\sqrt{g'(\tau)}}\frac{d}{d
\tau}\left(\frac{1}{\sqrt{f'(\tau)}}\right)-
\frac{1}{\sqrt{f'(\tau)}}\frac{d}{d
\tau}\left(\frac{1}{\sqrt{g'(\tau)}}\right)\right].
$$
After integrating, we
obtain
$$
\frac{1}{\sqrt{g'(\tau)}}\frac{d}{d
\tau}\left(\frac{1}{\sqrt{f'(\tau)}}\right)-
\frac{1}{\sqrt{f'(\tau)}}\frac{d}{d
\tau}\left(\frac{1}{\sqrt{g'(\tau)}}\right)=-c.\phantom{-----}(A.2)
$$
Then multiplying (A.2) by $f'(\tau)$ we arrive at the total
derivative
$$
\frac{f'(\tau)}{\sqrt{g'(\tau)}}\frac{d}{d
\tau}\left(\frac{1}{\sqrt{f'(\tau)}}\right)-
\sqrt{f'(\tau)}\frac{d}{d
\tau}\left(\frac{1}{\sqrt{g'(\tau)}}\right)=-\frac{d}{d\tau}
\sqrt{\frac{f'(\tau)}{g'(\tau)}}=-c f'(\tau).
$$
As a result, $g'(\tau)=f'(\tau)[c f(\tau)+d]^{-2}$, and thus,
$$
g(\tau)=\frac{a f(\tau)+b}{c f(\tau)+d},\quad\mbox{where}\;ad-bc=1.
$$

\end{document}